\documentclass[twocolumn,amsmath,amssymb]{revtex4}
\usepackage{graphicx}% Include figure files
\usepackage{dcolumn}% Align table columns on decimal point
\usepackage{bm}% bold math

\begin{document}

\preprint{APS/123-QED}

\title{Approximation scheme based on effective interactions for stochastic gene regulation}
\author{Jun Ohkubo}
\email[Email address: ]{ohkubo@i.kyoto-u.ac.jp}
\affiliation{
Graduate School of Informatics, Kyoto University,\\
36-1, Yoshida Hon-machi, Sakyo-ku, Kyoto-shi, Kyoto 606-8501, Japan
}
\date{\today}% It is always \today, today,

\begin{abstract}
Since gene regulatory systems contain sometimes only a small number of molecules,
these systems are not described well by macroscopic rate equations;
a master equation approach is needed for such cases.
We develop an approximation scheme for dealing with the stochasticity of the gene regulatory systems.
Using an effective interaction concept,
original master equations can be reduced to simpler master equations,
which can be solved analytically.
We apply the approximation scheme to self-regulating systems with monomer or dimer interactions,
and a two-gene system with an exclusive switch.
The approximation scheme can recover bistability of the exclusive switch adequately.
\end{abstract}

%\pacs{}% PACS, the Physics and Astronomy
\maketitle

\section{Introduction}

Recently, stochastic nature in small systems 
has attracted many attentions \cite{Rao2002,Elowitz2002,Kaern2005}.
One of the interesting examples of the stochasticity is a gene regulatory system;
it has been known experimentally that the gene regulatory systems show various phenomena
caused by intrinsic noise \cite{Gardner2000,Okano2008}.
The gene regulatory systems basically consists of genes, RNAs, and proteins.
The genes could sometimes be activated or repressed by regulatory proteins
known as transcription factors.
The number of regulatory proteins is sometimes very small, and there are large fluctuations.
From a theoretical point of view,
the gene regulatory systems have been studied a lot using Monte Carlo simulations 
(e.g., \cite{Schultz2007,Yoda2007}).
In addition, in order to gain insights into
mechanisms or functions of the gene regulatory systems,
many analytical studies have been done 
\cite{Hasty2000,Kepler2001,Bialek2001,Sasai2003,Hornos2005,Walczak2005,Kim2007,Ohkubo2007,Shahrezaei2008,Visco2009,Walczak2009}.
For example, if we consider a self-regulating gene with monomer binding interactions,
an exact solution has been already known \cite{Hornos2005}.
When one considers more complicated systems, some approximations are needed.
Such approximations have also been developed;
Fokker-Planck or Langevin equation approach \cite{Hasty2000,Kepler2001,Bialek2001},
a variational approach \cite{Sasai2003,Kim2007,Ohkubo2007},
and self-consistent proteomic field approximation \cite{Walczak2005}.

A gene regulatory system with only two genes and feedback mechanisms has been studied a lot
because it plays an important role as a genetic switch;
two distinct stable states emerges, 
and they could be switched either spontaneously or by external signals.
In mathematical description for the gene regulatory systems,
the RNAs are sometimes neglected for simplicity,
and only genes and regulatory proteins are considered.
When we construct a macroscopic rate equation, in which fluctuations in protein copy numbers
or gene expression states are neglected, 
the analysis for the rate equation tells us the following facts:
A system with two mutually repressing genes shows a bistability,
and cooperative binding of regulatory proteins is important for making the bistability 
\cite{Cherry2000,Warren2004}.
Here, the cooperative binding means that 
combinations of two or more proteins need to activate or repress genes.
The macroscopic rate equation gives multiple stable solutions,
and each solution corresponds to a stable state of the gene regulatory systems,
which causes the bistability.
Hence, for the cooperative binding cases,
it may be enough to use the macroscopic treatments in order to investigate
qualitative behavior of the bistability.
However, other studies have shown that 
a so-called exclusive switch
shows a bistability even when the macroscopic rate equations have only one solution 
\cite{Lipshtat2006,Loinger2007,Schultz2008}.
Although the bistability has been confirmed numerically using Monte Carlo simulations,
no exact or approximated analytical treatment has been proposed yet, to the best of our knowledge.

In the present paper, we develop a new approximation scheme for gene regulatory systems.
In the approximation scheme,
there is no need to use continuous description such as Fokker-Planck or Langevin equations,
and hence the smallness or discrete properties of the system are not neglected.
The basic idea of the approximation is similar to the ``self-consistent proteomic field approximation''
developed by Walczak \textit{et al.} \cite{Walczak2005}.
In the self-consistent proteomic field approximation,
a joint probability for all genes is approximated as a product of probability distributions for each gene,
and then the interactions between genes and regulatory proteins 
can be evaluated `exactly' in this approximation.
In \cite{Walczak2005}, only toggle switches, which consist of two genes, have been studied;
as denoted in the discussions in \cite{Walczak2005},
further approximation would be needed for self-regulating systems.
We here extend the concept of \cite{Walczak2005},
and develop a more applicable approximation scheme;
the interactions between genes and regulatory proteins are approximated firstly,
and an effective interaction is introduced.
The new approximation scheme
would be useful to treat more complicated cases, such as the exclusive switch.
The new approximation scheme enables us to give analytical expressions for 
probability distributions of the numbers of proteins, 
without loss of the discreteness property of the system.
The effective interactions are estimated self-consistently.
We will demonstrate the usefulness of the approximation scheme
by using self-regulating systems and the exclusive switch without cooperative interactions.
%Especially, we note that it is not possible to deal with
%the exclusive switch in the self-consistent proteomic field approximation,
%because the joint probability distribution cannot be expressed
%as a product of probability distributions for each gene in the exclusive switch case.

The present paper is constructed as follows.
In Sec.~II, we give a brief review of a stochastic model for gene regulation.
In Sec.~III, self-regulating systems are studied.
Section~\ref{sec_self_new_method} gives one of the important results in the present paper,
in which our approximation scheme is proposed.
The proposed approximation scheme 
is applied to the exclusive switch in Sec.~VI.
Section V is concluding remarks.

\section{Stochastic model for gene regulation}

We here briefly review the basic biology of genetic regulatory system
and a simplified stochastic model, for readers' convenience.

A gene regulatory system consists of many components, such as genes, RNAs, and proteins.
The transcription of a gene is initiated by a binding of RNA polymerase to 
a promoter site of the gene in the DNA.
The binding of regulatory proteins (or molecules), so-called transcription factors,
can sometimes regulate the transcription initiation.
These regulatory proteins bind to own target operator sites,
and they sometimes act as 
repressors (which repress the transcription) or activators (which enhance the transcription) 
of the transcription.
When the RNA polymerase binds to a gene,
the gene sequence is copied into a messenger RNA (mRNA),
and the mRNA is translated into a protein molecules by a ribosome enzyme complex.
The produced proteins are important to determine the phenotypic behavior of the cell.
In addition, regulation of transcription is one important way of controlling 
the phenotypic behavior,
and sometimes the produced proteins can become regulatory signals for genes.

Although all of the above reactions would be important for the gene regulatory systems,
the mRNA is sometimes neglected in stochastic modelling for simplicity.
That is, the translation from mRNAs to proteins are straightforward,
and then we assume that an activated gene directly increases the number of proteins.
In addition, we consider that a repressed gene cannot produce any proteins,
which makes analytical treatments much simpler \cite{Hornos2005}.

In the present paper, 
all regulatory proteins act as repressors.
If regulatory proteins are not binding to a gene, then
we call a state of the gene as `ON' state; if not, the gene is in `OFF' state.
A gene in OFF state cannot produce any proteins, as we assumed above.

\section{Self-regulating system}
\label{sec_self}

In this section, we will explain a new approximation scheme using a simplest model,
i.e., a self-regulating system.
Exact solutions for the self-regulating system with monomer interactions have already been known.
After reviewing the exact solutions,
we will propose a new approximation scheme.
The new approximation scheme will be applied to
the self-regulating systems with monomer and dimer interactions, respectively.

\subsection{Model}
\label{sec_self_model}

\begin{figure}
\begin{center}
  \includegraphics[width=40mm,keepaspectratio,clip]{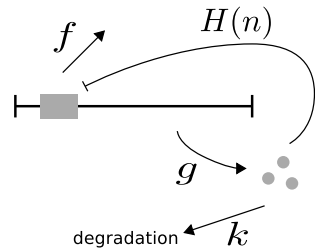}
\caption{
A schematic illustration of the self-regulating gene.
}
\label{fig_self_regulation}
\end{center}
\end{figure}

At first, we give a brief explanation for a self-regulating system.
In the self-regulating system,
there is only one gene, and it produces proteins.
The produced proteins are considered as regulatory proteins for the gene,
and the regulation is a repressed one.
In this sense, there is a self-regulation mechanism.
Figure~\ref{fig_self_regulation} shows the self-regulating system.
When the gene is in ON state, it produces proteins with rate $g$.
The degradation rate of the regulatory proteins is $k$.
The regulatory proteins can bind the gene with rate function $H(n)$,
where $n$ is the number of `free' regulatory proteins.
The function $H(n)$ can be a complicated function of the regulatory proteins;
e.g., $H(n) = h n$ for monomer interactions,
and $H(n) = h n (n-1)/2$ for dimer interactions,
where $h$ is a rate for the binding.
$f$ is the rate with which the regulatory protein is released from the repressor site of the gene.

\subsection{Exact solution for monomer interactions}
\label{sec_self_exact}

We here consider a simplest interaction case,
i.e., a monomer interaction case.
Hence, $\mathcal{H}(n)$ in Fig.~\ref{fig_self_regulation} is written as $h n$,
as discussed in Sec.~\ref{sec_self_model}.
For the monomer interaction cases,
exact solutions have already been known \cite{Hornos2005,Schultz2007}.
In order to compare our approximation scheme, which will be proposed in Sec.~\ref{sec_self_new_method},
we here briefly review the exact solutions.

\begin{figure}
\begin{center}
  \includegraphics[width=50mm,keepaspectratio,clip]{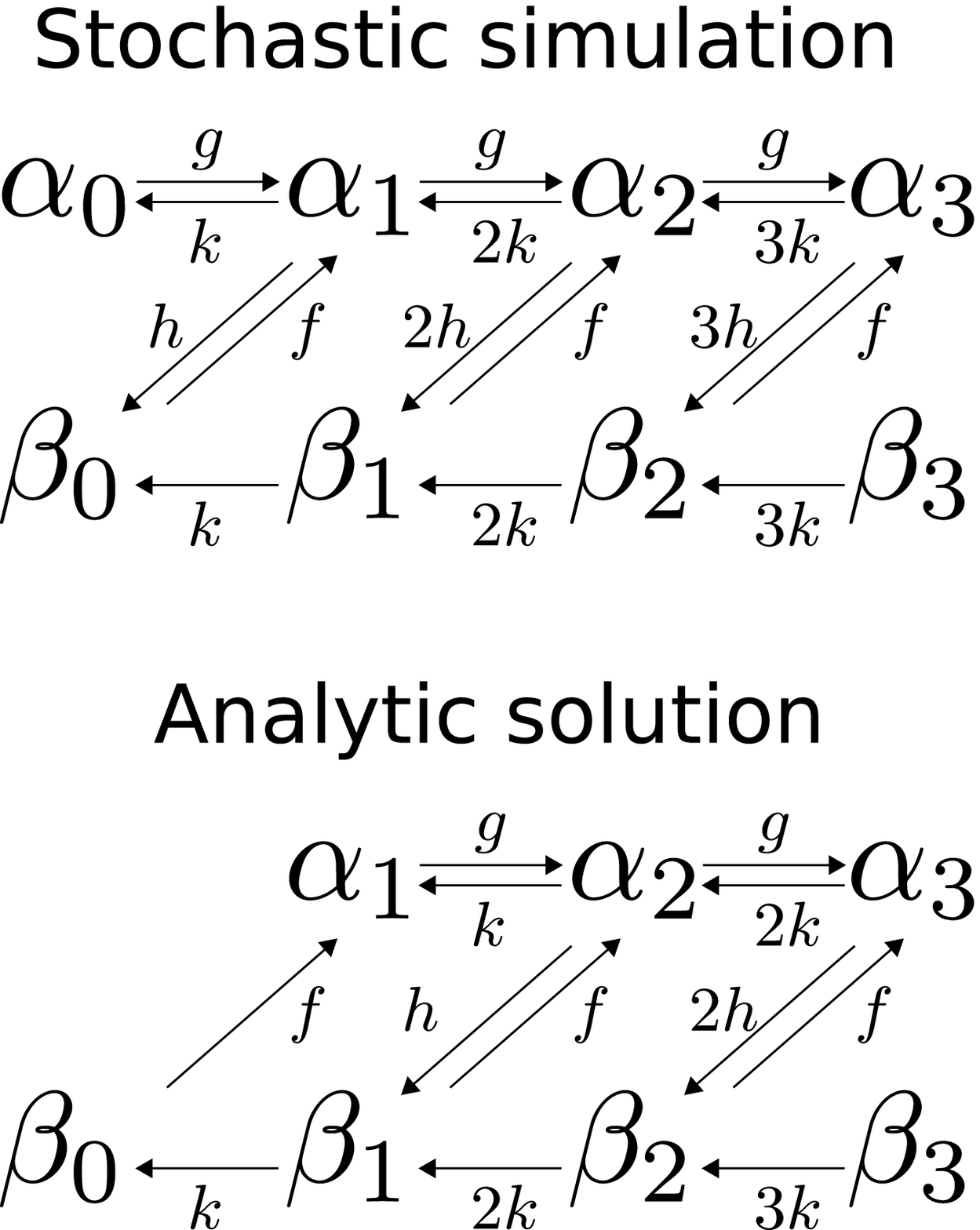}
\caption{
Transition scheme for simulations and analytical calculations
for monomer interaction cases.
$\alpha_n$ and $\beta_n$ correspond to probabilities with which
there are $n$ regulatory proteins for ON and OFF states, respectively.
In the analytical calculations, 
one of the proteins is considered as an inert one when the gene is in ON state,
and the inert protein is also included in $\alpha_n$;
the number of `free' regulatory proteins in ON state is $n-1$.
}
\label{fig_scheme_monomer}
\end{center}
\end{figure}

In order to make analytical treatments simpler,
one assumption should be included \cite{Schultz2007};
i.e., one of the proteins in ON state is inert,
and then the protein cannot be degraded or repress to the gene.
Hence, there are a little difference between usual stochastic simulations
and this analytical treatment.
However, it has already been discussed that
this assumption alter only for lower numbers of proteins,
and actually it gives quantitatively good results \cite{Schultz2007}.
Hence, we here employ this assumption.
Figure~\ref{fig_scheme_monomer} shows
the transition scheme for the usual stochastic simulations and the analytical treatment.
$\alpha_n$ and $\beta_n$ correspond to probabilities with which
there are $n$ regulatory proteins for ON and OFF states, respectively.
In the usual stochastic simulations,
the degradation rate of the proteins, i.e., the change
from $\alpha_n$ to $\alpha_{n-1}$,
is proportional to the number of proteins, $n$.
In contrast, 
the above assumption means that 
the rate from $\alpha_n$ to $\alpha_{n-1}$ in the analytical treatment
is proportional to $n-1$, not to $n$.

In this assumption,
the master equations are given as
\begin{align}
\frac{d \alpha'_n}{dt} =
& g [ \alpha'_{n-1} - \alpha'_n ]
+ k [(n+1)\alpha'_{n+1} - n \alpha'_n] \nonumber \\
&- h n \alpha'_n + f \beta_n, 
\label{eq_master_monomer_1}\\
\frac{d \beta_n}{dt} =
& + k [(n+1)\beta_{n+1} - n \beta_n]  \nonumber \\
&+ h n \alpha'_n - f \beta_n ,
\label{eq_master_monomer_2}
\end{align}
where $\alpha'_n$ 
is a probability with which 
there is $n$ `free' regulatory proteins for ON state;
$\alpha_{n+1} \equiv \alpha'_{n}$.
Note that the inert protein is not a `free' regulatory protein,
and $\alpha'_{n}$ does not include the inert protein.

In order to solve 
eqs.~\eqref{eq_master_monomer_1} and \eqref{eq_master_monomer_2},
it is useful to define the following generating functions;
\begin{align}
\alpha'(z) = \sum_{n=0}^\infty \alpha'_n z^n,\\
\beta(z) = \sum_{n=0}^\infty \beta_n z^n.
\end{align}
Using the generating functions,
various information about the self-regulating system can be obtained.
For example, 
the number of `free' regulatory proteins is 
\begin{align}
\alpha'_n = \left. \frac{\partial^n}{\partial z^n} \alpha'(z) \right|_{z=0}.
\end{align}
The probability with which the gene is in ON state
is given as $\alpha'(1)$;
the number of total proteins in the system is given as
\begin{align}
\langle n \rangle =
\left. \frac{\partial \alpha'(z)}{\partial z} \right|_{z=1}
+ 1 \times \alpha'(1)
+ \left. \frac{\partial \beta(z)}{\partial z} \right|_{z=1},
\label{eq_average}
\end{align}
where the second term in r.h.s means
a contribution from the inert protein in ON state.

Eqs.~\eqref{eq_master_monomer_1} and \eqref{eq_master_monomer_2}
can be rewritten as two differential equations
in terms of the generating functions;
\begin{align}
\frac{\partial \alpha(z)}{\partial t}
&= (z-1) \left[ g \alpha'(z) - k \frac{\partial \alpha'}{\partial z}
\right]
- h z \frac{\partial \alpha'}{\partial z}
+ f \beta(z), \\
\frac{\partial \beta(z)}{\partial t} 
&= - (z-1) k \frac{\partial \beta}{\partial z}
+ h z \frac{\partial \alpha'}{\partial z}
- f \beta(z).
\end{align}
After some calculations, 
`stationary' solutions for the generating functions
are obtained as follows \cite{Hornos2005,Schultz2007}:
\begin{align}
\alpha'^{\, \mathrm{ex}}(z) =& 
A^\mathrm{ex} \, F(1+a^\mathrm{ex}, 1+b^\mathrm{ex}, N^\mathrm{ex}(z-z_0^\mathrm{ex}) ), 
\label{eq_exact_alpha}\\
\beta^\mathrm{ex}(z) =& - \frac{1}{f}(z-1) \left[
g \alpha'^{\, \mathrm{ex}}(z) - k \frac{\partial \alpha'^{\, \mathrm{ex}}(z)}{\partial z}
\right] \nonumber \\
&+ \frac{h}{f} z \frac{\partial \alpha'^{\, \mathrm{ex}}(z)}{\partial z}
\label{eq_exact_beta},
\end{align}
where the super-script `$\mathrm{ex}$' means `exact solutions',
and
\begin{align}
&z^\mathrm{ex}_0 = \frac{k}{k+h}, \quad
N^\mathrm{ex} = \frac{g}{k+h}, \nonumber \\
&a^\mathrm{ex} = \frac{f}{k},  \quad
b^\mathrm{ex} = \frac{f}{k+h} + (1-z^\mathrm{ex}_0) N^\mathrm{ex}.
\end{align}
$A^\mathrm{ex}$ is the normalization constant,
which is determined as $\alpha'^{\, \mathrm{ex}}(1) + \beta^\mathrm{ex}(1) = 1$.
$F(p,q,r)$ is the Kummer confluent hypergeometric function,
\begin{align}
F(p,q,r) \equiv \sum_{n=0}^\infty \frac{(p)_n}{(q)_n} \frac{r^n}{n!},
\end{align}
where 
$(p)_n = p (p+1) (p+2) \cdots (p+n-1)$.

Details of the characteristics of the exact solution are written in \cite{Schultz2007}.
For example, 
the probability distributions for the numbers of `free' regulatory proteins are
\begin{align}
\alpha'^{\, \mathrm{ex}}_{n} =& \frac{A^\mathrm{ex}}{n!} (N^\mathrm{ex})^n 
\frac{(1+a^\mathrm{ex})_n}{(1+b^\mathrm{ex})_n} \nonumber \\
&\times F(1+a^\mathrm{ex}+n,1+b^\mathrm{ex}+n,-N^\mathrm{ex} z^\mathrm{ex}_0 ), \\
\beta^{\, \mathrm{ex}}_{n} =& \frac{A^\mathrm{ex}}{\tilde{f}} 
\left[ \left( (1-z_0^\mathrm{ex})N^\mathrm{ex} - b^\mathrm{ex} \right) \frac{(N^\mathrm{ex})^n}{n!}
\right. \nonumber \\
&\times \frac{(1+a^\mathrm{ex})_n}{(1+b^\mathrm{ex})_n} 
F(1+a^\mathrm{ex}+n,1+b^\mathrm{ex}+n,-N^\mathrm{ex} z^\mathrm{ex}_0) 
\nonumber \\
&\left. + b^\mathrm{ex} \frac{(N^\mathrm{ex})^n}{n!} \frac{(a^\mathrm{ex})_n}{(b^\mathrm{ex})_n}
F(a^\mathrm{ex}+n,b^\mathrm{ex}+n,-N^\mathrm{ex} z^\mathrm{ex}_0) \right],
\end{align}
where $\tilde{f} = f/ (k+h)$.

\subsection{Approximation scheme}
\label{sec_self_new_method}

In Sec.~\ref{sec_self_exact},
we analyzed the self-regulating system with monomer interactions.
In the system, the interaction factor, $H(n)$,
is given as $h n$,
and actually this simple form of the interaction
enables us to obtain the exact solutions.
If we consider different types of interactions, such as dimer ones,
exact solutions have not been known yet.

Here, we propose a new approximation scheme.
The key of the approximation scheme is to use ``an effective interaction''.
Although the new approximation scheme is similar to 
the self-consistent proteomic field approximation in \cite{Walczak2005},
the new one is more applicable;
it is applicable even to the self-regulating systems or exclusive switch cases,
as shown later.
%We here note that if we consider a toggle switch case discussed in \cite{Walczak2005},
%these approximations would give the same one.
The effective interaction means that the interaction factor in Fig.~\ref{fig_self_regulation}, $H(n)$,
is replaced as a scalar value;
i.e., $\mathcal{H}(n) = \tilde{h}$.
Here, the effective interaction $\tilde{h}$ is not a function of the regulatory proteins.
Hence, the master equation for this approximated system is written as follows:
\begin{align}
\frac{d \alpha'_n}{dt} =
& g [ \alpha'_{n-1} - \alpha'_n ]
+ k [(n+1)\alpha'_{n+1} - n \alpha'_n] \nonumber \\
&- \tilde{h} \alpha'_n + f \beta_n, 
\label{eq_master_monomer_1_app}\\
\frac{d \beta_n}{dt} =
& + k [(n+1)\beta_{n+1} - n \beta_n]  \nonumber \\
&+ \tilde{h} \alpha'_n - f \beta_n.
\label{eq_master_monomer_2_app}
\end{align}
Because the interaction factor $H(n) = \tilde{h}$ has a simple form,
the analytic solution can be easily calculated by using the generating function approach.
Putting the left-hand sides of Eqs.~\eqref{eq_master_monomer_1_app} and \eqref{eq_master_monomer_2_app}
as zero and rewriting Eqs.~\eqref{eq_master_monomer_1_app} and \eqref{eq_master_monomer_2_app}
in terms of the generating functions $\alpha'(z)$ and $\beta(z)$,
stationary solutions for the generating functions are obtained as follows:
\begin{align}
\alpha'(z) =& A \, F(a,b,N(z-1) ) 
\label{eq_app_alpha},\\
\beta(z) =& \left( 1 + \frac{h}{f} \right)
A \, F(a-1,b-1,N(z-1)) - \alpha(z)
\label{eq_app_beta},
\end{align}
where $A = f / (f+\tilde{h})$
and
\begin{align}
N = \frac{g}{k}, \quad a = 1 + \frac{f}{k}, \quad b = 1 + \frac{f+\tilde{h}}{k}.
\end{align}

A remaining task is to determine the effective interaction $\tilde{h}$.
For the self-regulating system with monomer interactions,
the binding of the regulatory proteins occurs only when the system is in ON state.
Hence, the number of proteins, which can be attached to the binding site, 
should be equal to the number of free proteins for ON state.

According to the following discussions,
we here set the effective interaction $\tilde{h}$ as
\begin{align}
\tilde{h} = h \langle n \rangle_{\alpha'},
\label{eq_effective_int_monomer}
\end{align}
where $\langle n \rangle_{\alpha'}$ is the expectation of the number of free regulatory proteins
under a condition that the gene is in ON state (conditional expectation).
Because it is possible to evaluate the conditional expectation using
the generating function (Eq.~\eqref{eq_app_alpha}) as
\begin{align}
\langle n \rangle_{\alpha'}
\equiv \frac{1}{\alpha'(1)} 
\left. \frac{\partial}{\partial z} \alpha'(z) \right|_{z=1}
= \frac{g(k+f)}{k(f+f+\tilde{h})},
\end{align}
we obtain the following self-consistent equation
by inserting Eq.~\eqref{eq_effective_int_monomer};
\begin{align}
h \langle n \rangle_{\alpha'}
= h \frac{g(k+f)}{k(f+f+h \langle n \rangle_{\alpha'})}.
\label{eq_monomer_self_consistent}
\end{align}
Solving Eq.~\eqref{eq_monomer_self_consistent},
we finally obtain 
\begin{align}
h \langle n \rangle_{\alpha'} =
\frac{-(k^2 + kf) + \sqrt{(k^2+kf)^2 + 3 k h g (k+f)}}{2k}.
\end{align}
Once the effective interaction $\tilde{h}$ is determined,
all statistical properties related to the number of regulatory proteins
are evaluated from the generating functions
(Eqs.~\eqref{eq_app_alpha} and \eqref{eq_app_beta}).
For example,
the probability distributions for the numbers of free proteins are
\begin{align}
\alpha'_n =& A N^n \frac{(a)_n}{(b)_n} F(a+n,b+n,-N), \\
\beta_n =& \left(1 + \frac{h}{f} \right) A N^n
\frac{(a-1)_n}{(b-1)_n} \nonumber \\
&\times F(a-1+n,b-1+n,-N) - \alpha'_n.
\end{align}

\subsection{Results for monomer interactions}
\label{sec_result_monomer}

\begin{figure}
\begin{center}
  \includegraphics[width=80mm,keepaspectratio,clip]{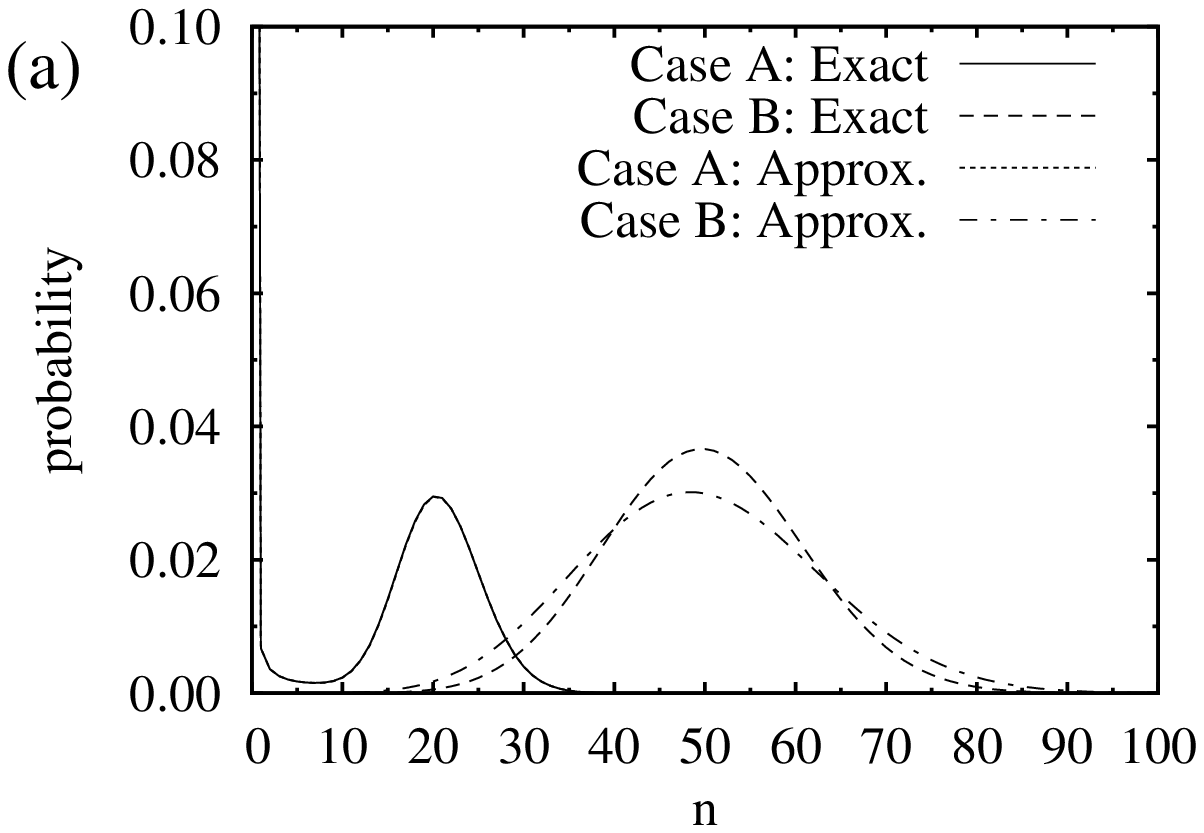}\\
  \includegraphics[width=80mm,keepaspectratio,clip]{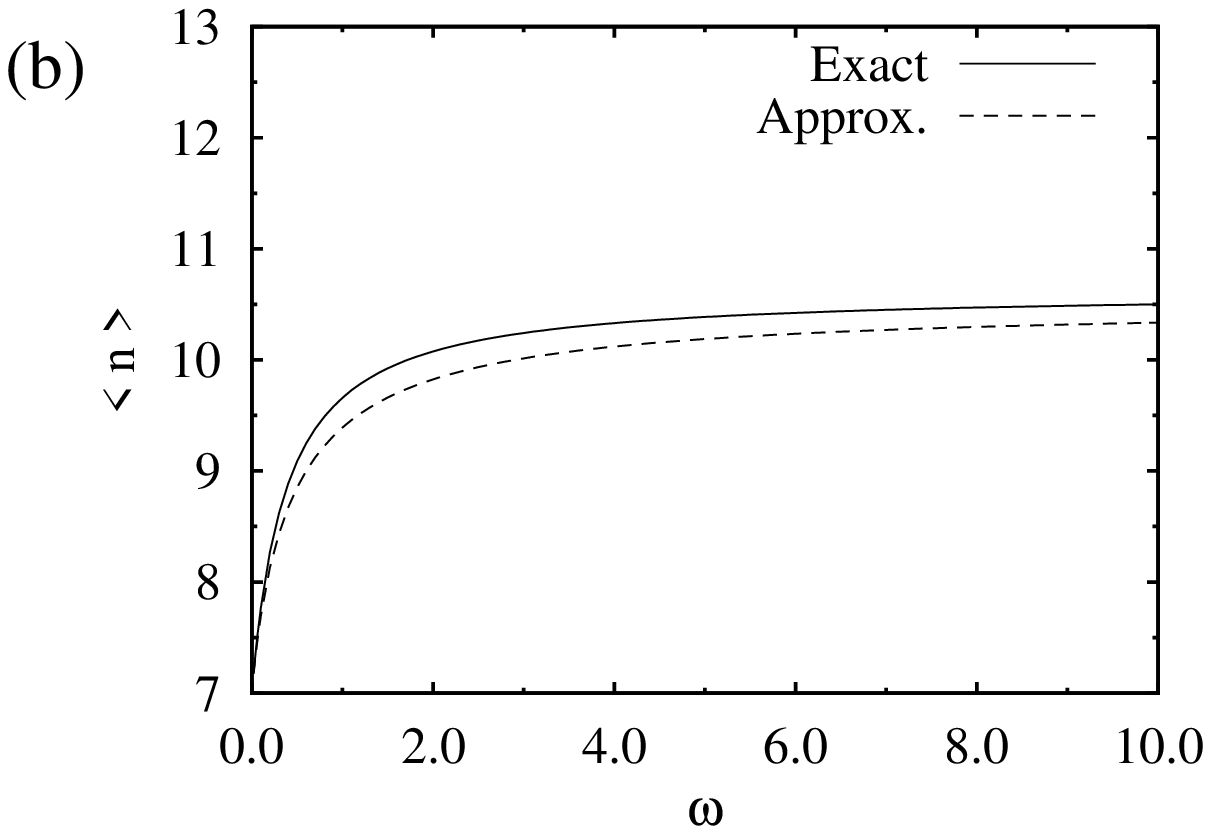}\\
\caption{
Comparison between results from the exact solutions and 
the approximation scheme.
(a) The probability distributions of the number of proteins. 
Case A: $X_\mathrm{eq} = 10.0$, $X_\mathrm{ad} = 10.0$, $\omega = 0.01$.
Case B: $X_\mathrm{eq} = 50.0$, $X_\mathrm{ad} = 50.0$, $\omega = 10.0$.
For Case A, 
it is difficult to see the difference
between the exact solution and approximate solution.
(b) The average number of proteins.
$X_\mathrm{eq} = 10.0$ and $X_\mathrm{ad} = 10.0$,
and only the value of $\omega$ was changed.
}
\label{fig_result_monomer}
\end{center}
\end{figure}

For monomer interaction cases, the exact solutions are obtained.
Hence, we here compare the exact results and approximate results
obtained by the approximation scheme.

For the comparison, we here introduce rescaled parameters
as follows \cite{Schultz2007}:
\begin{align}
\omega = \frac{f}{k}, 
\quad X_\mathrm{eq} = \frac{f}{h},
\quad X_\mathrm{ad} = \frac{g}{2k},
\end{align}
and for simplicity, we set $k = 1$ in all numerical evaluations.
These rescaled parameters are helpful to understand
properties of the genetic switch.
The parameter $X_\mathrm{ad}$ characterizes
the synthesis/degradation processes,
and large $X_\mathrm{ad}$ would give 
a large average number of proteins.
$X_\mathrm{eq}$ is related to the equilibrium constant
of the binding/unbinding process.
Finally, $\omega$ is a parameter called ``adiabaticity parameter''.
$\omega$ measures how rapidly the gene can equilibrate in a gene state.
If $\omega$ is small,
the synthesis/degradation behaves almost like an independent 
birth and death process,
and there would be two peaks corresponding to the binding/unbinding states,
respectively.
For details of these parameters,
e.g., see \cite{Schultz2007}.

Firstly, the probability distributions of the number of protein
were compared.
Figure~\ref{fig_result_monomer}(a) shows the results.
Here, we performed two cases:
In Case A, we set $X_\mathrm{eq} = 10.0$, $X_\mathrm{ad} = 10.0$, $\omega = 0.01$;
in Case B, $X_\mathrm{eq} = 50.0$, $X_\mathrm{ad} = 50.0$, $\omega = 10.0$.
For Case A, the exact solution and the approximate one
give a good agreement, and it is difficult to see the difference.
Although there are quantitative differences between the exact and approximate solutions
for Case B,
the approximation scheme gives qualitatively good result
despite of the rough approximation.
Figure~\ref{fig_result_monomer}(b) shows 
the average number of proteins (Eq.~\eqref{eq_average}) for various values of $\omega$
when $X_\mathrm{eq} = 10.0$ and $X_\mathrm{ad} = 10.0$.
If $\omega$ is small, the approximation scheme
gives quantitatively good results.
Even in the large $\omega$ case,
the difference between the exact and approximate results
are less than 1.

\subsection{Results for dimer interactions}

\begin{figure}
\begin{center}
  \includegraphics[width=60mm,keepaspectratio,clip]{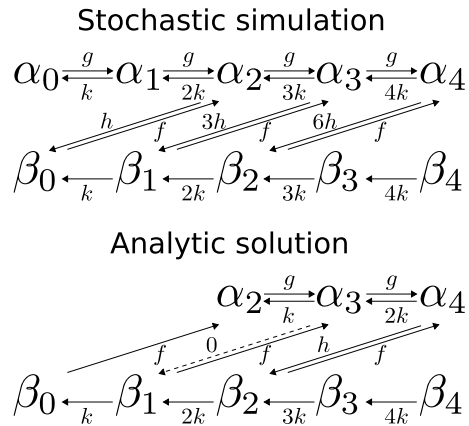}
\caption{
Transition scheme for simulations and analytical calculations
for dimer interaction cases.
}
\label{fig_scheme_dimer}
\end{center}
\end{figure}

As a second example, a self-regulating system
with dimer interactions are considered.
In this case, the transition scheme for analytical calculations
are different from the monomer interaction cases;
see Fig.~\ref{fig_scheme_dimer}.
In this case, the master equations are
the same as Eqs.~\eqref{eq_master_monomer_1_app} and \eqref{eq_master_monomer_2_app},
but the effective interaction should be set as
\begin{align}
\tilde{h} = h \frac{\langle n (n-1) \rangle_{\alpha'}}{2},
\end{align}
and we should interpret $\alpha'_n$ as
$\alpha_{n+2} = \alpha'_n$.

Using the similar procedure written in Sec.~\ref{sec_self_new_method},
the effective interaction $\tilde{h}$ is obtained
by solving the following self-consistent equation:
\begin{align}
h \frac{\langle n (n-1) \rangle_{\alpha'}}{2}
= \frac{h}{2} \frac{1}{\alpha'(1)} 
\left. \frac{\partial^2}{\partial z^2} \alpha'(z) \right|_{z=0}.
\label{eq_dimer_self_consistent}
\end{align}
Since it is a little complicated task to obtain the analytical expression
for the effective interaction $\tilde{h}$,
we numerically solved the self-consistent equation
(Eq.~\eqref{eq_dimer_self_consistent}).

\begin{figure}
\begin{center}
  \includegraphics[width=80mm,keepaspectratio,clip]{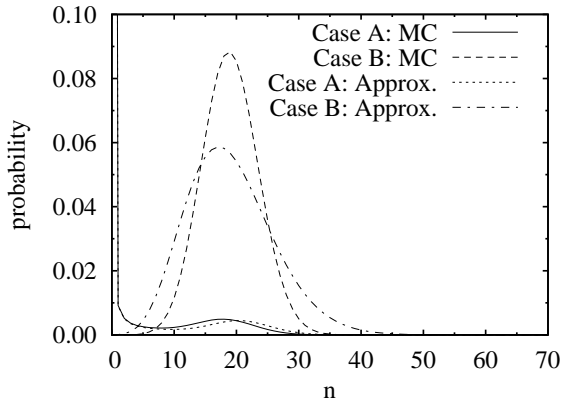}
\caption{
The probability distributions of the number of proteins.
Case A: $X_\mathrm{eq} = 10.0$, $X_\mathrm{ad} = 10.0$, $\omega = 0.01$.
Case B: $X_\mathrm{eq} = 50.0$, $X_\mathrm{ad} = 50.0$, $\omega = 10.0$.
}
\label{fig_result_dimer}
\end{center}
\end{figure}

For the dimer interaction cases,
we have not obtained any exact solution.
Hence, results of the approximation scheme
were compared with those of the Monte Carlo simulations.

Figure~\ref{fig_result_dimer} shows the probability distributions.
In Case A, we used the following rescaled parameters:
$X_\mathrm{eq} = 10.0$, $X_\mathrm{ad} = 10.0$, $\omega = 0.01$;
in Case B, $X_\mathrm{eq} = 50.0$, $X_\mathrm{ad} = 50.0$, $\omega = 10.0$.
The numbers of the Monte Carlo steps are over $10^7$ for Case A,
and $10^8$ for Case B.
Obviously, the approximation scheme gives qualitatively good results;
although the shapes of the distributions and the positions of peaks
are slightly different,
the number of peaks are the same as the Monte Carlo results.
In addition, the average number of proteins are almost the same
as the Monte Carlo results:
for Case A, $\langle n \rangle = 1.2$ in the Monte Carlo simulation
and $\langle n \rangle = 1.3$ in the approximation scheme;
for Case B, $\langle n \rangle = 19.0$ in the Monte Carlo simulation
and $\langle n \rangle = 19.0$ in the approximation scheme.

\section{Exclusive switch}

\begin{figure}
\begin{center}
  \includegraphics[width=60mm,keepaspectratio,clip]{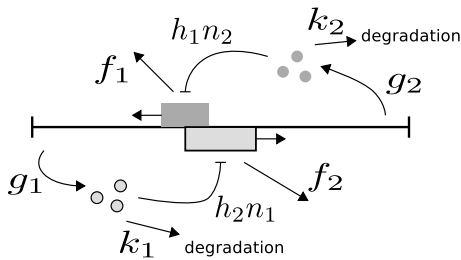}
\caption{
A schematic illustration of the exclusive switch.
}
\label{fig_result}
\end{center}
\end{figure}

Next, we consider a more complicated case, i.e., an exclusive switch 
\cite{Lipshtat2006,Loinger2007,Schultz2008}.
The exclusive switch consists of two genes, gene 1 and gene 2.
The two genes have overlapping promoter sites,
and the binding of one of the regulatory proteins prevents the binding 
of the other regulatory proteins.
%Such a situation is seen in nature, e.g., in $\lambda$-phage regulation.

Here, the interaction between the binding sites and proteins is assumed to be monomer interactions.
Because the interaction is not cooperative bindings,
the macroscopic rate equation gives only one solution \cite{Lipshtat2006,Loinger2007,Schultz2008}.
Although there is only one solution,
it has been shown that the exclusive switch can play as a switch.
In the exclusive switch, 
stochastic effects make the bistability even without cooperativity
between the regulatory proteins.

In order to study the exclusive switch analytically,
master equations for a joint probability $P(n_1, n_2, s_1, s_2)$ should be constructed;
$n_i \in \bm{N}$ is the number of proteins for gene $i$,
and $s_i \in \{ \mathrm{ON}, \mathrm{OFF} \}$ indicates the gene state.
In general, the master equations
for multiple gene cases are very complicated,
and it could be difficult to obtain numerical solutions
if the number of genes is large.

However, in our approximation scheme, the effective interaction is used,
and it enables us to reduce the equations to be solved.
Because we consider only the effective interaction,
the probability $P(n_1, n_2, s_1, s_2)$
can be expressed as $P(n_1, s_1) \times P(n_2, s_2)$.
For example, the master equation for gene 1 is written as
\begin{align}
\frac{d \alpha'^{(1)}_n}{dt} =
& g^{(1)} [ \alpha'^{(1)}_{n-1} - \alpha'^{(1)}_n ]
+ k^{(1)} [(n+1)\alpha'^{(1)}_{n+1} - n \alpha'^{(1)}_n] \nonumber \\
&- \tilde{h}^{(1)}\alpha'^{(1)}_n + f^{(1)} \beta^{(1)}_n, \\
\frac{d \beta^{(1)}_n}{dt} =
& + k^{(1)} [(n+1)\beta^{(1)}_{n+1} - n \beta^{(1)}_n]  \nonumber \\
&+ \tilde{h}^{(1)} \alpha'^{(1)}_n - f^{(1)} \beta^{(1)}_n,
\end{align}
where the super-script `$(1)$' indicates gene 1.
Master equations for gene 2 can be obtained in the similar way.
Using the same discussion in the self-regulating systems in Sec.~\ref{sec_self},
the generating functions for genes $i \in \{1,2\}$, $\alpha'^{(i)}(z)$ and $\beta^{(i)}(z)$,
are derived.

The effective interaction $\tilde{h}^{(1)}$ should be chosen as follows.
The transition of gene 1 from ON state to OFF state
can occur only when the gene 2 is in ON state,
and the effective interaction $\tilde{h}^{(1)}$ includes
only a contribution from the free proteins 2 in ON state.
Note that gene 1 does not know whether gene 2 is in ON state or OFF state,
different from the self-regulating system discussed in Sec.~\ref{sec_self};
in the self-regulating system, the gene knows the own state.
Hence, the evaluation of the effective interaction is slightly different
from the self-regulating systems.
The conditional average of the number of free proteins 2
is given by
$\langle n^{(2)} \rangle_{\alpha'} = (\partial \alpha'^{(2)}(z) / \partial z |_{z=1}) 
/ \alpha'^{(2)}(1)$,
as discussed in Sec.~\ref{sec_self_new_method}.
The probability $P(2_\mathrm{ON})$, with which gene 2 is in ON state,
is calculated as $P(2_\mathrm{ON}) = \alpha'^{(2)}(1)$.
Defining
$\langle n^{(2)} \rangle'_{\alpha'} \equiv \langle n^{(2)} \rangle_{\alpha'} \, P(2_\mathrm{ON})$,
the effective interaction should be written as
\begin{align}
\tilde{h}^{(1)} = h^{(1)} \langle n^{(2)} \rangle'_{\alpha'}.
\end{align}
According to the above discussions,
we finally obtain the following self-consistent equations;
\begin{align}
&h^{(2)} \langle n^{(1)} \rangle'_{\alpha'} = \nonumber \\
&\quad \frac{h^{(2)}  g^{(1)} f^{(1)} (k^{(1)} + f^{(1)})}
{
k^{(1)} ( f^{(1)} + h^{(1)} \langle n^{(2)} \rangle'_{\alpha'} ) 
(k^{(1)} + f^{(1)} + h^{(1)} \langle n^{(2)} \rangle'_{\alpha'} ) 
}, \\
&h^{(1)} \langle n^{(2)} \rangle'_{\alpha'} = \nonumber \\
&\quad \frac{h^{(1)}  g^{(2)} f^{(2)} (k^{(2)} + f^{(2)})}
{
k^{(2)} ( f^{(2)} + h^{(2)} \langle n^{(1)} \rangle'_{\alpha'} ) 
(k^{(2)} + f^{(2)} + h^{(2)} \langle n^{(1)} \rangle'_{\alpha'} ) 
}.
\end{align}
By solving the above self-consistent equations,
we obtain $\tilde{h}^{(1)}$ and $\tilde{h}^{(2)}$.
We here solve them numerically.

\begin{figure*}
\begin{center}
  \includegraphics[width=70mm,keepaspectratio,clip]{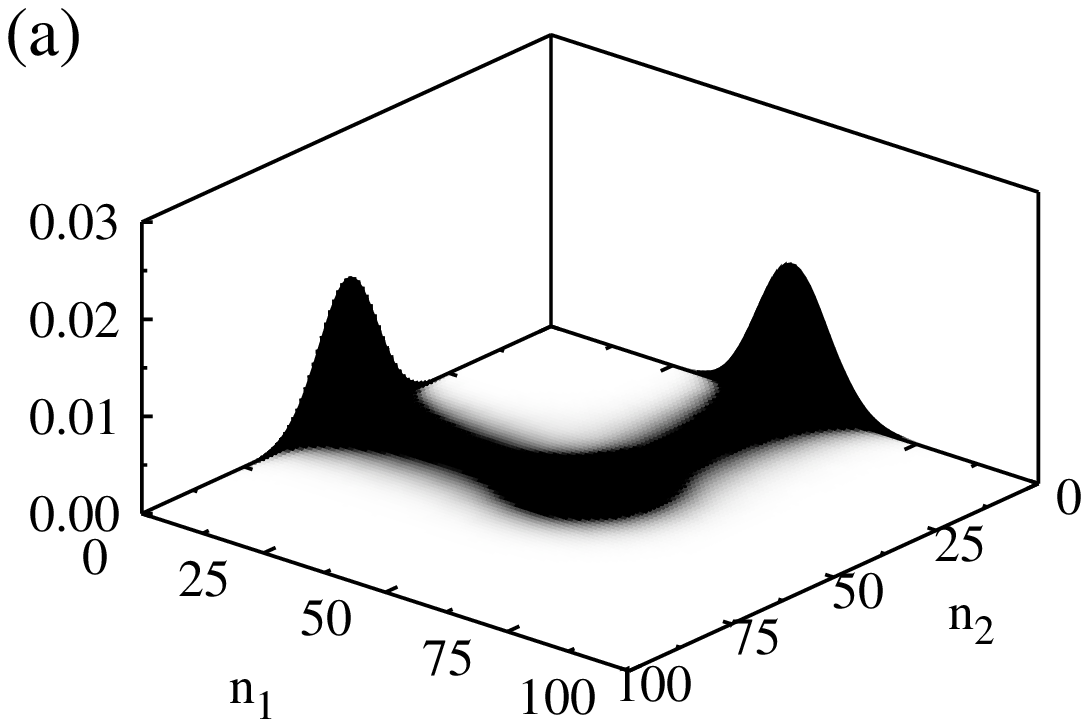}
  \includegraphics[width=70mm,keepaspectratio,clip]{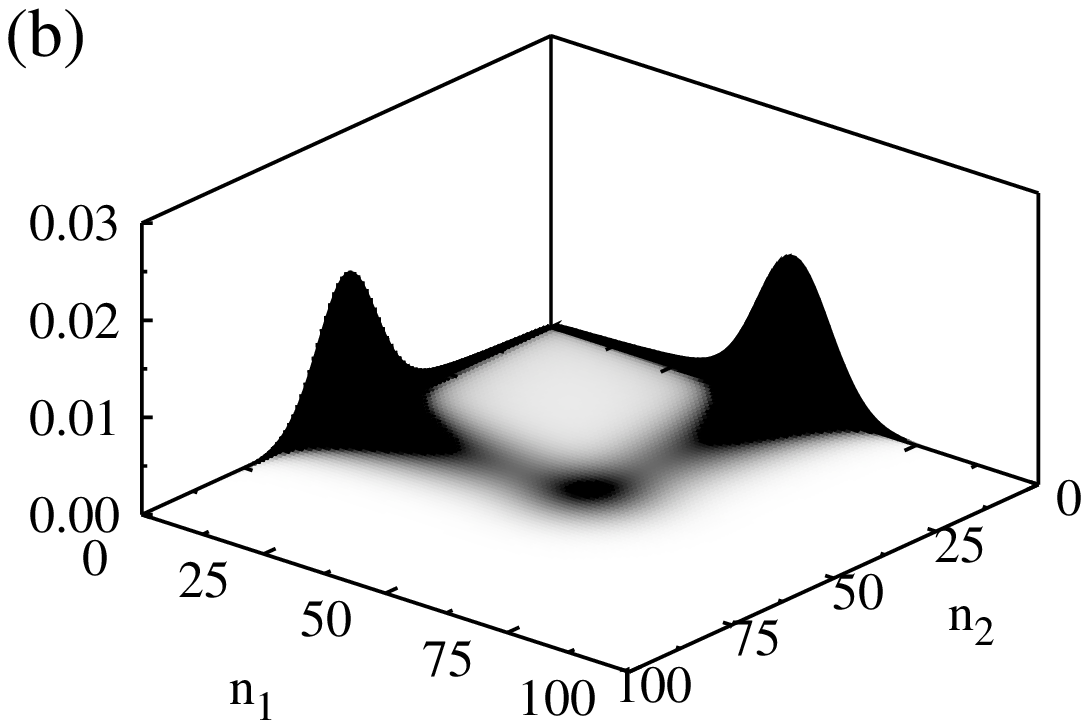} \\
  \includegraphics[width=70mm,keepaspectratio,clip]{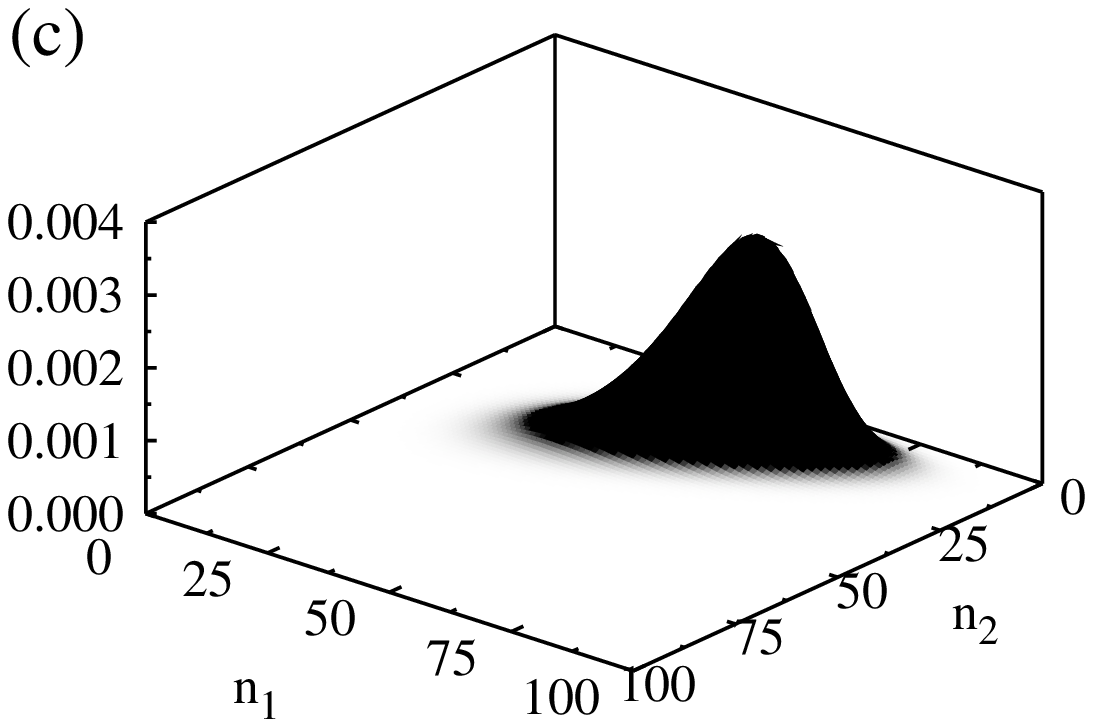}
  \includegraphics[width=70mm,keepaspectratio,clip]{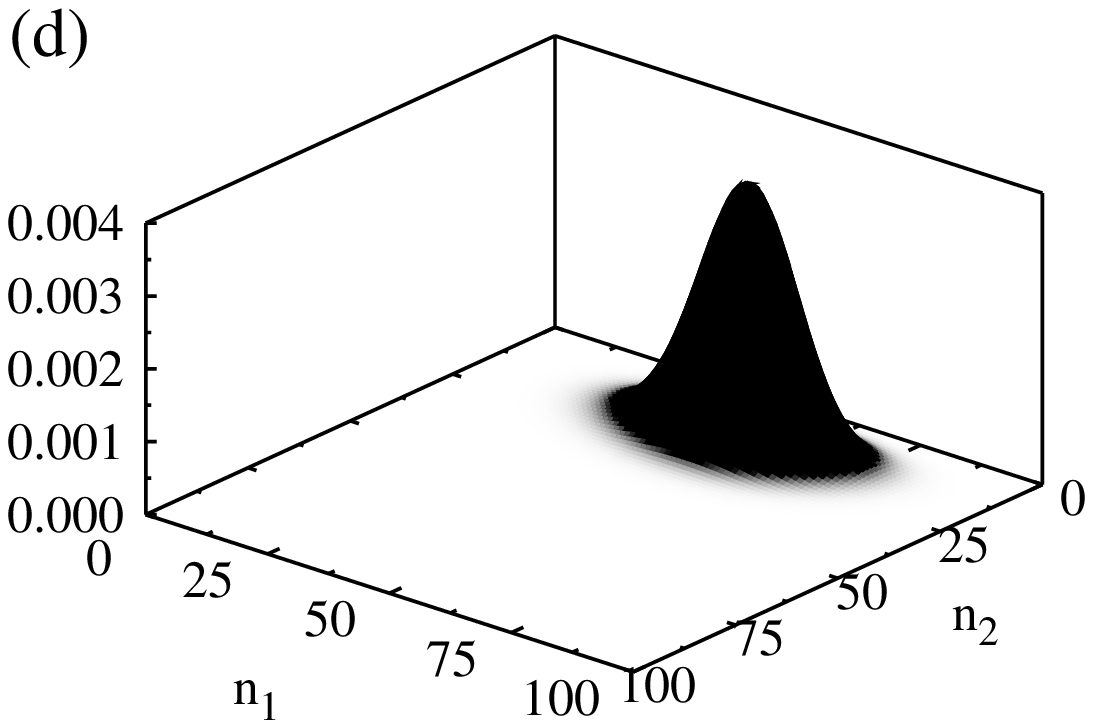} 
\caption{
Joint probability distributions for the exclusive switch.
(a) Monte Carlo results for $X^{(1)}_\mathrm{eq} = X^{(2)}_\mathrm{eq} = 25.0$, 
$X^{(1)}_\mathrm{ad} = X^{(2)}_\mathrm{ad} = 25.0$,
and $\omega^{(1)} = \omega^{(2)} = 0.1$.
(b) Approximate results.
All parameters are the same as (a).
(c) Monte Carlo results for 
$X^{(1)}_\mathrm{eq} = 40.0$, $X^{(2)}_\mathrm{eq} = 30.0$, 
$X^{(1)}_\mathrm{ad} = 30.0$, $X^{(2)}_\mathrm{ad} = 20.0$,
$\omega^{(1)} = 10.0$ and $\omega^{(2)} = 20.0$.
(d) Approximate results.
All parameters are the same as (c).
}
\label{fig_result_exclusive}
\end{center}
\end{figure*}

We can immediately obtain the probability distributions for each gene
using the approximation scheme.
In order to reconstruct the joint probability distribution for genes 1 and 2,
we need more calculations as follows.
Firstly, we calculate conditional probabilities for the number of free proteins
for gene $i$ ($i \in \{1,2\}$) as
\begin{align}
\tilde{\alpha}'^{(i)}_n \equiv \frac{\alpha'^{(i)}_{n}}{\alpha'^{(i)}(1)}, \\
\tilde{\beta}^{(i)}_n \equiv \frac{\beta^{(i)}_{n}}{\beta^{(i)}(1)}. 
\end{align}
Secondly, because of the approximation scheme,
%the probability distributions for each gene can be
%considered as a kind of marginal distributions,
the joint probability distribution
should be evaluated as
\begin{align}
P(n_1, n_2, 1_\mathrm{ON}, 2_\mathrm{ON})
&= P(1_\mathrm{ON}, 2_\mathrm{ON}) \tilde{\alpha}'^{(1)}_{n_1-1} \tilde{\alpha}'^{(2)}_{n_2-1}, \\
P(n_1, n_2, 1_\mathrm{ON}, 2_\mathrm{OFF})
&= P(1_\mathrm{ON}, 2_\mathrm{OF}) \tilde{\alpha}'^{(1)}_{n_1-1} \tilde{\beta}^{(2)}_{n_2}, \\
P(n_1, n_2, 1_\mathrm{OFF}, 2_\mathrm{ON})
&= P(1_\mathrm{OFF}, 2_\mathrm{ON}) \tilde{\beta}^{(1)}_{n_1} \tilde{\alpha}'^{(2)}_{n_2-1},
\end{align}
where $P(1_\mathrm{ON},2_\mathrm{ON})$ is the probability
with which gene 1 is in ON state and gene 2 is in ON state, and so on.
Note that $\alpha'$ means only the number of `free' proteins;
for monomer interaction cases, the difference between $\alpha$ and $\alpha'$ 
is only one inert protein.
In addition, the probability with which both genes 1 and 2 are in OFF state is zero;
$P(1_\mathrm{OFF},2_\mathrm{OFF}) = 0$,
because of the exclusive settings.

Taking the exclusive settings into account,
the marginal probabilities are calculated as follows: 
\begin{align}
\begin{cases}
P(1_\mathrm{ON}, 2_\mathrm{ON} ) + P(1_\mathrm{OFF}, 2_\mathrm{ON})
= P(2_\mathrm{ON} ), \\
P(1_\mathrm{ON},2_\mathrm{OFF}) = P(2_\mathrm{OFF}),\\
P(1_\mathrm{OFF},2_\mathrm{ON}) = P(1_\mathrm{OFF}),\\
P(1_\mathrm{ON}, 2_\mathrm{ON} ) + P(1_\mathrm{ON}, 2_\mathrm{OFF})
= P(1_\mathrm{ON} ) ,
\end{cases}
\end{align}
and then
\begin{align}
\begin{cases}
P(1_\mathrm{ON}, 2_\mathrm{ON} ) = P(1_\mathrm{ON}) - P(2_\mathrm{OFF}), \\
P(1_\mathrm{ON}, 2_\mathrm{OFF}) = P(2_\mathrm{OFF}), \\
P(1_\mathrm{OFF}, 2_\mathrm{ON}) = P(1_\mathrm{OFF}), \\
P(1_\mathrm{OFF}, 2_\mathrm{OFF}) = 0.
\end{cases}
\end{align}
The marginal probabilities, such as $P(1_\mathrm{ON})$,
can be evaluated by using the generating function $\alpha'^{(i)}(z)$ and $\beta^{(i)}(z)$.
Finally, we can construct the joint probability distribution as 
\begin{align}
P(n_1, n_2) =& P(n_1,n_2,1_\mathrm{ON},2_\mathrm{ON}) \nonumber \\
&+ P(n_1,n_2,1_\mathrm{OFF},2_\mathrm{ON}) \nonumber \\
&+ P(n_1,n_2,1_\mathrm{ON},2_\mathrm{OFF}).
\end{align}
We here note that 
the probabilities 
$P(1_\mathrm{ON},2_\mathrm{ON})$,
calculated using the above procedures,
may become negative for some cases;
i.e., $P(1_\mathrm{ON}) > P(2_\mathrm{OFF})$ for some choices of parameters 
$g^{(i)}, k^{(i)}, h^{(i)}$ and $f^{(i)}$.
In these cases,
other procedures to estimate the joint probabilities are needed.
In the following numerical experiments,
only the former cases ($P(1_\mathrm{ON}) < P(2_\mathrm{OFF})$) are treated.
%the joint probability would be calculated using different methods;
%e.g., we set 
%\begin{align}
%P(s_1, s_2) = 
%\begin{cases}
%0 & \textrm{($s_1 = $ OFF, $s_2 = $ OFF)}, \\
%\frac{P(s_1) P(s_2)}{P_\mathrm{total}} & \textrm{(otherwise)},
%\end{cases}
%\end{align}
%where
%\begin{align}
%P_\mathrm{total} = &
%P(1_\mathrm{ON}) P(2_\mathrm{ON})
%+ P(1_\mathrm{ON}) P(2_\mathrm{OFF}) \nonumber \\
%&+ P(1_\mathrm{OFF}) P(2_\mathrm{ON}).
%\end{align}

Figure~\ref{fig_result_exclusive} shows the joint probability distributions.
Figures~\ref{fig_result_exclusive}(a) and (c) are Monte Carlo results,
and Figs.~\ref{fig_result_exclusive}(b) and (d) are results of the approximation scheme.
As in Sec.~\ref{sec_result_monomer},
we used the rescaled parameters,
and set $k^{(1)} = k^{(2)} = 1$.
In Figs.~\ref{fig_result_exclusive}(a) and (b),
we used the parameters
$X^{(1)}_\mathrm{eq} = X^{(2)}_\mathrm{eq} = 25.0$, 
$X^{(1)}_\mathrm{ad} = X^{(2)}_\mathrm{ad} = 25.0$,
and $\omega^{(1)} = \omega^{(2)} = 0.1$;
for (c) and (d),
$X^{(1)}_\mathrm{eq} = 40.0$, $X^{(2)}_\mathrm{eq} = 30.0$, 
$X^{(1)}_\mathrm{ad} = 30.0$, $X^{(2)}_\mathrm{ad} = 20.0$,
$\omega^{(1)} = 10.0$ and $\omega^{(2)} = 20.0$.
The numbers of the Monte Carlo steps are over $10^8$ for Fig.~\ref{fig_result_exclusive}(a),
and over $10^9$ for Fig.~\ref{fig_result_exclusive}(c).
Although Fig.~\ref{fig_result_exclusive}(d) does not show
the correlated behavior seen in Fig.~\ref{fig_result_exclusive}(c)
because correlations between gene 1 and gene 2 are largely neglected
in the approximation scheme,
one could say that the approximation scheme gives qualitatively good results;
the characteristics of the peak structure can be recovered adequately
despite the rough approximation.
In Fig.~\ref{fig_result_exclusive}(b), 
the bistability due to the exclusive settings is recovered well.

%The average number of proteins are as follows:
%\begin{itemize}
%\item[(a)] $\langle n^{(1)} \rangle = 12.0$, $\langle n^{(2)} \rangle = 12.0$
%\item[(b)] $\langle n^{(1)} \rangle = 10.5$, $\langle n^{(2)} \rangle = 10.5$
%\item[(c)] $\langle n^{(1)} \rangle = 51.1$, $\langle n^{(2)} \rangle = 18.5$
%\item[(d)] $\langle n^{(1)} \rangle = 52.3$, $\langle n^{(2)} \rangle = 16.5$
%\end{itemize}

\section{Concluding remarks}

In the present paper, we developed the approximation scheme for gene regulatory systems.
We firstly applied it to self-regulating systems.
The approximation scheme gives qualitatively good results;
the characteristics of peak structures can be recovered well.
In addition, due to the extension of the basic idea of the effective interactions,
we can naturally apply the approximation scheme even to the exclusive switch,
and the bistability of the exclusive switch without cooperative interactions is successfully recovered.

In contrast to the Fokker-Planck or Langevin approach,
the approximation scheme proposed in the present paper does not neglect 
discrete properties of systems.
In addition, because we can rewritten the joint probability for all genes
as a product of probability distribution for each gene,
the dimensions of the problems are reduced largely.

The approximation scheme developed in the present paper
would be a very crude one;
it cannot treat correlated characteristics between genes.
However, approximate analytic expressions are immediately obtained,
and qualitatively good results are given despite the crude approximation.
Since Monte Carlo simulations are sometimes time-consuming
and need high computational costs,
it would be beneficial to study such approximation scheme
in order to obtain qualitative pictures for the probability distributions.
In addition, developments of analytical treatments
would be helpful to gain insights and understandings for the regulatory systems

\section*{ACKNOWLEDGMENTS}

The author thank Masaki Sasai for helpful comments for this manuscript.
This work was supported in part by grant-in-aid for scientific research 
(Nos. 20115009 and 21740283)
from the Ministry of Education, Culture, Sports, Science and Technology (MEXT), Japan.

\end{document}